\documentclass[11pt,a4paper]{article}
\usepackage{amsfonts}
\usepackage{amsmath,amssymb}
\usepackage{epsfig,graphicx}

\begin{document}

\begin{center}
{\Huge Lorentzian Goldstone modes }

{\Huge \bigskip shared among photons and gravitons}

{\Huge \bigskip }

\bigskip

\bigskip

\textbf{\ J.L.~Chkareuli, J. Jejelava and Z. Kepuladze}

\bigskip

\textit{Center for Elementary Particle Physics, Ilia State University, 0162
Tbilisi, Georgia\ \vspace{0pt}\\[0pt]
}

\textit{and} \textit{E. Andronikashvili} \textit{Institute of Physics, 0177
Tbilisi, Georgia\ }

\bigskip \bigskip \bigskip \bigskip \bigskip

\textbf{Abstract}

\bigskip
\end{center}

It has long been known that photons and gravitons may appear as vector and
tensor Goldstone modes caused \ by spontaneous Lorentz invariance violation
(SLIV). Usually this approach is considered for photons and gravitons
separately. We develop the emergent electrogravity theory consisting of the
ordinary QED and the tensor field gravity model which mimics the linearized
general relativity in Minkowski spacetime. In this theory, Lorentz symmetry
appears incorporated into higher global symmetries of the length-fixing
constraints put on the vector and tensor fields involved, $A_{\mu }^{2}=\pm
M_{A}^{2}$ and $H_{\mu \nu }^{2}=\pm M_{H}^{2}$ ($M_{A}$ and $M_{H}$ are the
proposed symmetry breaking scales). We show that such a SLIV pattern being
related to breaking of global symmetries underlying these constraints
induces the massless Goldstone and pseudo-Goldstone modes shared among
photon and graviton. While for a vector field case the symmetry of the
constraint coincides with Lorentz symmetry $SO(1,3)$ of the electrogravity
Lagrangian, the tensor field constraint itself possesses much higher global
symmetry $SO(7,3)$, whose spontaneous violation provides a sufficient number
of zero modes collected in a graviton. Accordingly, while photon may only
contain true Goldstone modes, graviton appears at least partially composed
from pseudo-Goldstone modes rather than from pure Goldstone ones. When
expressed in terms of these modes, the theory looks essentially nonlinear
and contains a variety of Lorentz and $CPT$ violating couplings. However,
all SLIV effects turn out to be strictly cancelled in the lowest order
processes that is considered in some detail. How this emergent
electrogravity theory could be observationally differed from conventional
QED and GR theories is also briefly discussed.

\thispagestyle{empty}\newpage

\section{Introduction}

An extremely successful concept of the spontaneously broken internal
symmetries in particle physics allows to think that a spontaneous violation
of spacetime symmetries and, particularly, a spontaneous Lorentz invariance
violation (SLIV), could also provide some dynamical approach to quantum
electrodynamics \cite{bjorken}, gravity \cite{ohan} and Yang-Mills theories
\cite{eg} with photon, graviton and non-Abelian gauge fields appearing as
massless Nambu-Goldstone (NG) bosons \cite{NJL} (for some later
developments, see \cite{cfn,jb,kraus,jen,bluhm,kan,kos,car}). In this
connection, we recently suggested \cite{cjt} an alternative approach to the
emergent gravity theory in the framework of nonlinearly realized Lorentz
symmetry for the underlying symmetric two-index tensor field in a theory,
which mimics linearized general relativity in Minkowski space-time. It was
shown that such a SLIV pattern, due to which a true vacuum in the theory is
chosen, induces massless tensor Goldstone and pseudo-Goldstone modes some of
which can naturally be associated with the physical graviton.

This approach itself has had a long history, dating back to the model of
Nambu \cite{nambu} for QED with a nonlinearly realized Lorentz symmetry for
the underlying vector field. This may indeed appear through the
"length-fixing" vector field constraint

\begin{equation}
A_{\mu }^{2}=n^{2}M_{A}^{2}\text{ , \ \ }A_{\mu }^{2}\equiv A_{\mu }A^{\mu }%
\text{, }\ n^{2}\equiv n_{\nu }n^{\nu }=\pm 1  \label{const}
\end{equation}%
(where $n_{\nu }$ is a properly oriented unit Lorentz vector, while $M_{A}$
is the proposed scale for Lorentz violation) much as it works in the
nonlinear $\sigma $-model \cite{GL} for pions, $\sigma ^{2}+\pi ^{2}=f_{\pi
}^{2}$, where $f_{\pi }$ is the pion decay constant. Note that a
correspondence with the nonlinear $\sigma $ model for pions may appear
rather suggestive in view of the fact that pions are the only presently
known Goldstone particles whose theory, chiral dynamics\cite{GL}, is given
by the nonlinearly realized chiral $SU(2)\times SU(2)$ symmetry rather than
by an ordinary linear $\sigma $ model. The constraint (\ref{const}) means in
essence that the vector field $A_{\mu }$ develops some constant background
value

\begin{equation}
<A_{\mu }(x)>\text{ }=n_{\mu }M_{A}  \label{vev1}
\end{equation}%
and the Lorentz symmetry $SO(1,3)$ formally breaks down to $SO(3)$ or $%
SO(1,2)$ depending on the timelike ($n^{2}>0$) or space-like ($n^{2}<0$)
nature of SLIV. However, in sharp contrast to the nonlinear $\sigma $ model
for pions, the nonlinear QED theory, due to the starting gauge invariance
involved, ensures that all the physical Lorentz violating effects turn out
to be non-observable. It was shown \cite{nambu}, while only in the tree
approximation and for the timelike SLIV ($n^{2}>0$), that the nonlinear
constraint (\ref{const}) implemented into the standard QED Lagrangian
containing a charged\ fermion field $\psi (x)$

\begin{equation}
L_{QED}=-\frac{1}{4}F_{\mu \nu }F^{\mu \nu }+\overline{\psi }(i\gamma
\partial +m)\psi -eA_{\mu }\overline{\psi }\gamma ^{\mu }\psi \text{ \ \ }
\label{lag11}
\end{equation}%
as a supplementary condition appears in fact as a possible gauge choice for
the vector field $A_{\mu }$, while the $S$-matrix remains unaltered under
such a gauge convention. Really, this nonlinear QED contains a plethora of
Lorentz and $CPT$ violating couplings when it is expressed in terms of the
pure emergent photon modes ($a_{\mu }$) according to the constraint
condition (\ref{const})

\begin{equation}
A_{\mu }=a_{\mu }+n_{\mu }(M_{A}^{2}-n^{2}a^{2})^{\frac{1}{2}}\text{ , \ }%
n_{\mu }a_{\mu }=0\text{ \ \ \ \ (}a^{2}\equiv a_{\mu }a^{\mu }\text{).}
\label{gol}
\end{equation}%
For definiteness, one takes the positive sign for the square root (giving an
effective Higgs mode) when expanding it in powers of $a^{2}/M_{A}^{2}$
\begin{equation}
A_{\mu }=a_{\mu }+M_{A}n_{\mu }-\frac{n^{2}}{2M_{A}}a^{2}n_{\mu
}+O(1/M_{A}^{2})  \label{constr}
\end{equation}%
However, the contributions of all these Lorentz violating couplings to
physical processes completely cancel out among themselves. So, SLIV is shown
to be superficial as it affects only the gauge of the vector potential $%
A_{\mu }$ at least in the tree approximation \cite{nambu}.

Some time ago, this result was extended to the one-loop approximation and
for both\ the timelike ($n^{2}>0$) and space-like ($n^{2}<0$) Lorentz
violation \cite{az}. All the contributions to the photon-photon,
photon-fermion and fermion-fermion interactions violating physical Lorentz
invariance was shown to exactly cancel among themselves in the manner
observed long ago by Nambu for the simplest tree-order diagrams. This means
that the constraint (\ref{const}), having been treated as a nonlinear gauge
choice at the tree (classical) level, remains as a gauge condition when
quantum effects are taken into account as well. So, in accordance with
Nambu's original conjecture, one can conclude that physical Lorentz
invariance is left intact at least in the one-loop approximation, provided
we consider the standard gauge invariant QED Lagrangian (\ref{lag11}) taken
in flat Minkowski space-time. Later this result was also confirmed for the
spontaneously broken massive QED \cite{kep}, non-Abelian theories \cite{jej}
and tensor field gravity \cite{cjt}. The point is, however, that all these
calculations represent somewhat "empirical" confirmation of gauge invariance
of the nonlinear QED and other emergent theories rather than the theoretical
one. Indeed, whether the constraint (\ref{const}) amounts in general to a
special gauge choice for a vector field is an open question unless the
corresponding gauge function satisfying the constraint condition is
explicitly constructed. We discuss this important issue in more detail in
section 4.

Let us note that, in principle, the vector field constraint (\ref{const})
may be formally obtained in some limit from a convential potential that
could be included in the QED Lagrangian (\ref{lag11})%
\begin{equation}
U(A)=\lambda _{A}(A_{\mu }^{2}-n^{2}M_{A}^{2})^{2}  \label{u1}
\end{equation}%
thus extending QED to the so-called bumblebee model \cite{bum}. \ Here $%
\lambda _{A}>0$ stands for the coupling constant of the vector field, while
values of $n^{2}=\pm 1$ determine again its possible vacuum configurations.
Indeed, one can readily see that the potential (\ref{u1}) inevitably causes
spontaneous violation of Lorentz symmetry in an ordinary way, much as an
internal symmetry violation is caused in the linear $\sigma $ model for
pions \cite{GL}. As a result, one has a massive \textquotedblleft Higgs"
mode (with mass $2\sqrt{2\lambda _{A}}M_{A}$) together with massless
Goldstone modes associated with the photon components. However, as was
argued in \cite{kkk}, the bumblebee model adding the potential terms (\ref%
{u1}) to the standard QED Lagrangian is generally unstable. Indeed, its
Hamiltonian appears unbounded from below unless the phase space is
constrained just by the nonlinear condition $A_{\mu }^{2}=n^{2}M_{A}^{2}$.
With this condition imposed, the Hamiltonian becomes positive, the massive
Higgs mode never emerges, and the model is physically equivalent to the
Nambu model \cite{nambu}. Remarkably, this pure Goldstone theory limit can
be reached when, just as in the $\sigma $ model for pions, one goes from the
linear model for the SLIV to the nonlinear one by taking the limit $\lambda
_{A}\rightarrow \infty $. This immediately fixes in (\ref{u1}) the vector
field square to its vacuum value thus leading to the above constraint (\ref%
{const}). As matter of fact, the vector field theory turns out to be stable
in this limit only.

Actually, for the tensor field gravity we use the similar nonlinear
constraint for a symmetric two-index tensor field
\begin{equation}
H_{\mu \nu }^{2}=\mathfrak{n}^{2}M_{H}^{2}\text{ , \ }H_{\mu \nu }^{2}\equiv
H_{\mu \nu }H^{\mu \nu }\text{, \ }\mathfrak{n}^{2}\equiv \mathfrak{n}_{\mu
\nu }\mathfrak{n}^{\mu \nu }=\pm 1\text{ }  \label{const3}
\end{equation}%
(where $\mathfrak{n}_{\mu \nu }$ is now a properly oriented unit Lorentz
tensor, while $M_{H}$ is the proposed scale for Lorentz violation in the
gravity sector) which fixes its length in the same manner as it appears for
the vector field (\ref{const}). Again, the nonlinear constraint (\ref{const3}%
) may in principle appear from the standard potential terms added to the
tensor field Lagrangian%
\begin{equation}
U(H)=\lambda _{H}(H_{\mu \nu }^{2}-\mathfrak{n}^{2}M_{H}^{2})^{2}\text{ }
\label{u11}
\end{equation}%
in the nonlinear $\sigma $-model type limit when the coupling constant $%
\lambda _{H}$ goes to infinity. Just in this limit the tensor field theory
appears stable, though, due to the corresponding Higgs mode excluded, it
does not lead to physical Lorentz violation \cite{cjt}.

Usually, an emergent gauge field framework is considered either regarding
emergent photons or regarding emergent gravitons. For the first time, we
consider it regarding them both in the so-called electrogravity theory where
together with the Nambu QED model \cite{nambu} with its gauge invariant
Lagrangian (\ref{lag11}) we propose the linearized Einstein-Hilbert kinetic
term for the tensor field preserving a diffeomorphism (diff) invariance. We
show that such a combined SLIV pattern, conditioned by the constraints (\ref%
{const}) and (\ref{const3}), induces the massless Goldstone modes which
appear shared among photon and graviton. Note that one needs in common nine
zero modes both for photon (three modes) and graviton (six modes) to provide
all necessary (physical and auxiliary) degrees of freedom. They actually
appear in our electrogravity theory due to spontaneous breaking of high
symmetries of the constraints involved. While for a vector field case the
symmetry of the constraint coincides with Lorentz symmetry $SO(1,3)$, the
tensor field constraint itself possesses much higher global symmetry $%
SO(7,3) $, whose spontaneous violation provides a sufficient number of zero
modes collected in a graviton. These modes are largely pseudo-Goldstone
modes (PGMs) since $SO(7,3)$ is symmetry of the constraint (\ref{const3})
rather than the electrogravity Lagrangian whose symmetry is only given by
Lorentz invariance. The electrogravity theory we start with becomes
essentially nonlinear, when expressed in terms of the Goldstone modes, and
contains a variety of \ Lorentz (and $CPT$) violating couplings. However, as
our calculations show, all SLIV effects turn out to be strictly cancelled in
the low order physical processes involved once the tensor field gravity part
of the electrogravity theory is properly extended to general relativity
(GR). This can be taken as an indication that in the electrogravity theory
physical Lorentz invariance is preserved in this approximation. Thereby, the
length-fixing constraints (\ref{const}, \ref{const3})\ put on the vector and
tensor fields appear as the gauge fixing conditions rather than sources of
the actual Lorentz violation just as it was in the pure nonlinear QED
framework \cite{nambu}. From this viewpoint, if this cancellation were to
work in all orders, one could propose that emergent theories, like as the
electrogravity theory, are not differed from conventional gauge theories. We
argue, however, that even in this case some observational difference between
them could unavoidably appear, if gauge invariance were presumably broken by
quantum gravity at the Planck scale order distances.

The paper is organized in the following way. In section 2 we formulate the
model for the tensor field gravity and find corresponding massless Goldstone
modes some of which are collected in the graviton. Then in section 3 we
consider in significant detail the combined electrogravity theory consisting
of QED and tensor field gravity. In the subsequent section 4 we derive
general Feynman rules for basic interactions in the emergent framework. The
model appears in essence three-parametric containing the inverse Planck and
SLIV scales, $1/M_{P}$, $1/M_{A}$ and $1/M_{H},$ respectively, as the
perturbation parameters, so that the SLIV interactions are always
proportional some powers of them. Further, some lowest order SLIV processes,
such as an elastic photon-graviton scattering and photon-graviton conversion
are considered in detail. We show that all these effects, taken in the tree
approximation, appear in fact vanishing so that the physical Lorentz
invariance is ultimately restored. Finally, in section 5 we present our
conclusion.

\section{Tensor field gravity}

We propose here, closely following our earlier papers \cite{cjt}, the tensor
field gravity theory which mimics linearized general relativity in Minkowski
space-time. The corresponding Lagrangian for one real vector field $A_{\mu }$
(still representing all sorts of matter in the model)

\begin{equation}
\mathcal{L}(H,A)=\mathcal{L}(H)+\mathcal{L}(A)+\mathcal{L}_{int}  \label{tl}
\end{equation}%
consists of the tensor field kinetic\ terms of the form

\begin{equation}
\mathcal{L}(H)=\frac{1}{2}\partial _{\lambda }H^{\mu \nu }\partial ^{\lambda
}H_{\mu \nu }-\frac{1}{2}\partial _{\lambda }H_{tr}\partial ^{\lambda
}H_{tr}-\partial _{\lambda }H^{\lambda \nu }\partial ^{\mu }H_{\mu \nu
}+\partial ^{\nu }H_{tr}\partial ^{\mu }H_{\mu \nu }\text{ ,}  \label{fp}
\end{equation}%
($H_{tr}$ stands for the trace of $H_{\mu \nu },$ $H_{tr}=\eta ^{\mu \nu
}H_{\mu \nu }$), which is invariant under the diff transformations

\begin{equation}
\delta H_{\mu \nu }=\partial _{\mu }\xi _{\nu }+\partial _{\nu }\xi _{\mu }%
\text{ , \ \ \ }\delta x^{\mu }=\xi ^{\mu }(x)\text{ ,}  \label{tr3}
\end{equation}%
and the interaction terms

\begin{equation}
\text{ \ \ \ \ }\mathcal{L}_{int}(H,A)=-\frac{1}{M_{P}}H_{\mu \nu }T^{\mu
\nu }(A)\text{ . \ \ \ \ \ }  \label{fh}
\end{equation}%
The $\mathcal{L}(A)$ and $T^{\mu \nu }(A)$ are the conventional free
Lagrangian and energy-momentum tensor for a vector field

\begin{equation}
\mathcal{L}(A)=-\frac{1}{4}F_{\mu \nu }F^{\mu \nu },\text{ }T^{\mu \nu
}(A)=-F^{\mu \rho }F_{\rho }^{\nu }+\frac{1}{4}\eta ^{\mu \nu }F_{\alpha
\beta }F^{\alpha \beta }  \label{tt}
\end{equation}%
It is clear that, in contrast to the tensor field kinetic \ terms, the other
terms in (\ref{tl}) are only approximately invariant under the diff
transformations (\ref{tr3}). They become more and more invariant when the
tensor field gravity Lagrangian (\ref{tl}) is properly extended to GR with
higher terms in $H$-fields included\footnote{%
Such an extension means that in all terms included in the GR action,
particularly in the\ QED Lagrangian term , $(-g)^{1/2}g_{\mu \nu }g_{\lambda
\rho }F^{\mu \lambda }F^{\nu \rho }$, one expands the metric tensors
\begin{equation*}
g_{\mu \nu }=\eta _{\mu \nu }+H_{\mu \nu }/M_{P}\text{, \ }g^{\mu \nu }=\eta
^{\mu \nu }-H^{\mu \nu }/M_{P}+H^{\mu \lambda }H_{\lambda }^{\nu
}/M_{P}^{2}+\cdot \cdot \cdot
\end{equation*}%
taking into account the higher terms in $H$-fields.}. Following the
nonlinear $\sigma $-model for QED \cite{nambu}, we propose the SLIV
condition (\ref{const3}) as some tensor field length-fixing constraint which
is supposed to be substituted into the total Lagrangian $\mathcal{L}(H,A)$
prior to the variation of the action. This eliminates, as was mentioned
above, a massive Higgs mode in the final theory thus leaving only massless
Goldstone modes, some of which are then collected in a graviton.

Let us first turn to the spontaneous Lorentz violation itself \ in a gravity
sector, which is caused by the constraint (\ref{const3}), while such a
violation in a QED sector is assumed to be determined by the constraint (\ref%
{const}). The latter leads, as was mentioned above, only\ to two possible
breaking channels of the starting Lorentz symmetry, namely to $SO(3)$ or $%
SO(1,2)$ depending on the timelike ($\mathfrak{n}^{2}>0$) or space-like ($%
\mathfrak{n}^{2}<0$) nature of SLIV. For the tensor field constraint (\ref%
{const3}) the choice appears wider. Indeed, this constraint can be written
in the more explicit form

\begin{equation}
H_{\mu \nu }^{2}=H_{00}^{2}+H_{i=j}^{2}+(\sqrt{2}H_{i\neq j})^{2}-(\sqrt{2}%
H_{0i})^{2}=\mathfrak{n}^{2}M_{H}^{2}=\pm \text{ }M_{H}^{2}\text{ \ \ \ \ \
\ \ }  \label{c4}
\end{equation}%
(where the summation over all indices $(i,j=1,2,3)$ is imposed) and means in
essence that the tensor field $H_{\mu \nu }$ develops the vacuum expectation
value (VEV) configuration

\begin{equation}
<H_{\mu \nu }(x)>\text{ }=\mathfrak{n}_{\mu \nu }M_{H}  \label{v}
\end{equation}%
determined by the matrix $\mathfrak{n}_{\mu \nu }$. The initial Lorentz
symmetry $SO(1,3)$ of the Lagrangian $\mathcal{L}(H,A)$ given in (\ref{tl})
then formally breaks down at a scale $M_{H}$ to one of its subgroups. If one
assumes a "minimal" vacuum configuration in the $SO(1,3)$ space with the VEV
(\ref{v}) developed on a single $H_{\mu \nu }$ component, there are in fact
the following three breaking channels

\begin{eqnarray}
(a)\text{ \ \ \ \ }\mathfrak{n}_{00} &\neq &0\text{ , \ \ }%
SO(1,3)\rightarrow SO(3)  \notag \\
(b)\text{ \ \ \ }\mathfrak{n}_{i=j} &\neq &0\text{ , \ \ }SO(1,3)\rightarrow
SO(1,2)  \label{ns} \\
(c)\text{ \ \ \ }\mathfrak{n}_{i\neq j} &\neq &0\text{ , \ \ }%
SO(1,3)\rightarrow SO(1,1)  \notag
\end{eqnarray}%
for the positive sign in (\ref{c4}), and

\begin{equation}
(d)\text{ \ \ }\mathfrak{n}_{0i}\neq 0\text{ , \ \ }SO(1,3)\rightarrow SO(2)
\label{nss}
\end{equation}%
for the negative sign. These cases can be readily derived taking an
appropriate exponential parametrization for the tensor field
\begin{equation}
H_{\alpha \beta }=\left[ e^{i\mathfrak{n}_{\mu \nu }\mathcal{J}^{\mu \sigma
}\eta ^{\nu \tau }h_{\sigma \tau }/M_{H}}\right] _{\alpha \beta }^{\gamma
\delta }\mathfrak{n}_{\gamma \delta }M_{H}\text{, \ \ }\left( \mathcal{J}%
^{\mu \nu }\right) _{\alpha }^{\beta }=i\left( \delta _{\alpha }^{\mu }\eta
^{\nu \beta }-\delta _{\alpha }^{\nu }\eta ^{\mu \beta }\right)  \label{np}
\end{equation}%
where the Lorenz generators $\mathcal{J}^{\mu \nu }$ are explicitly included%
\footnote{%
One may alternatively argue starting from the vector representation of the
higher $SO(7,3)\mathfrak{\ }$symmetry\ determined by the constraint equation
(\ref{c4}) itself (see below). Thereby, one has a standard parametrization
\begin{equation*}
H_{A}=\left[ e^{i\mathfrak{n}_{M}\mathcal{J}^{MN}h_{N}/M_{H}}\right] _{A}^{B}%
\mathfrak{n}_{B}M_{H}
\end{equation*}%
where the "big" indices $A,B,M,N$ correspond to the pairs of different
values of the old indices ($\mu \nu $) appeared in (\ref{c4}). Consequently,
one has an equality $\mathfrak{n}_{M}\mathcal{J}^{MN}h_{N}=\mathfrak{n}_{\mu
\nu }\mathcal{J}^{\mu \sigma }\eta ^{\nu \tau }h_{\sigma \tau }$ when going
to the standard Lorentz indices so that antisymmetry in the indices ($M,N$)
goes to antisymmetry in the index pairs ($\mu \nu ,\sigma \tau $).}.
Accordingly, there are only three Goldstone modes in the cases ($a,b$) and
five modes in the cases ($c$-$d$). In order to associate at least one of the
two transverse polarization states of the physical graviton with these
modes, one could have any of the above-mentioned SLIV channels except for
the case ($a$) where the only nonzero Goldstone modes are given by the
tensor components $h_{0i}$ ($i=1,2,3$). Indeed, it is impossible for
graviton to have all vanishing spatial components, as appears in the case ($%
a $). However, these components may be provided by some accompanying
pseudo-Goldstone modes, as we argue below. Apart from the minimal VEV
configuration, there are many others as well. A particular case of interest
is that of the traceless VEV tensor $\mathfrak{n}_{\mu \nu }$

\begin{equation}
\text{\ \ }\mathfrak{n}_{\mu \nu }\eta ^{\mu \nu }=0  \label{tll}
\end{equation}%
in terms of which the emergent gravity Lagrangian acquires an especially
simple form (see below). It is clear that the VEV in this case can be
developed on several $H_{\mu \nu }$ components simultaneously, which in
general may lead to total Lorentz violation with all six Goldstone modes
generated. For simplicity, we will use sometimes this form of vacuum
configuration in what follows, while our arguments can be applied to any
type of VEV tensor $\mathfrak{n}_{\mu \nu }$.

Aside from the pure Lorentz Goldstone modes, the question of the other
components of the symmetric two-index tensor $H_{\mu \nu }$ naturally
arises. Remarkably, they turn out to be pseudo-Goldstone modes (PGMs) in the
theory. Indeed, although we only propose Lorentz invariance of the
Lagrangian $\mathcal{L}(H,A)$, the SLIV constraint (\ref{const3}) formally
possesses the much higher global accidental symmetry $SO(7,3)$ of the
constrained bilinear form (\ref{c4}), which manifests itself when
considering the $H_{\mu \nu }$ components as the "vector" ones under $%
SO(7,3) $. This symmetry is in fact spontaneously broken, side by side with
Lorentz symmetry, at the scale $M_{H}.$ Assuming again a minimal vacuum
configuration in the $SO(7,3)$ \ space, with the VEV (\ref{v}) developed on
a single $H_{\mu \nu }$ component, we have either timelike ($SO(7,3)$ $%
\rightarrow SO(6,3)$) or space-like ($SO(7,3)$ $\rightarrow SO(7,2)$)
violations of the accidental symmetry depending on the sign of $\mathfrak{n}%
^{2}=\pm 1$ in (\ref{c4}). According to the number of broken $SO(7,3)$
generators, just nine massless Goldstone modes appear in both cases.
Together with an effective Higgs component, on which the VEV is developed,
they complete the whole ten-component symmetric tensor field $H_{\mu \nu }$
of the basic Lorentz group as is presented in its parametrization (\ref{par}%
). Some of them are true Goldstone modes of the spontaneous Lorentz
violation, others are PGMs since, as was mentioned, an accidental $SO(7,3)$
symmetry is not shared by the whole Lagrangian $\mathcal{L}(H,A)$ given in (%
\ref{tl}). Notably, in contrast to the scalar PGM case \cite{GL}, they
remain strictly massless being protected by the starting diff invariance
which becomes exact when the tensor field gravity Lagrangian (\ref{tl}) is
properly extended to GR$^{1}$. Owing to this invariance, some of the Lorentz
Goldstone modes and PGMs can then be gauged away from the theory, as usual.

Now, one can rewrite the Lagrangian $\mathcal{L}(H,A)$ in terms of the
tensor Goldstone modes explicitly using the SLIV constraint (\ref{const3}).
For this purpose, let us take the following handy parameterization for the
tensor field $H_{\mu \nu }$

\begin{equation}
H_{\mu \nu }=h_{\mu \nu }+\mathfrak{n}_{\mu \nu }(M_{H}^{2}-\mathfrak{n}%
^{2}h^{2})^{\frac{1}{2}},\text{ }\mathfrak{n}\cdot h=0\text{\ \ \ }(%
\mathfrak{n}\cdot h\equiv \mathfrak{n}_{\mu \nu }h^{\mu \nu })\text{ .}
\label{par}
\end{equation}%
Here $h_{\mu \nu }$ corresponds to the pure emergent modes satisfying the
orthogonality condition, while the effective \textquotedblleft Higgs" mode
(or the $H_{\mu \nu }$ component in the vacuum direction) is given by the
square root for which we takes again the positive sign when expanding it in
powers of $h^{2}/M_{H}^{2}$ ($h^{2}\equiv h_{\mu \nu }h^{\mu \nu }$)
\begin{equation}
\text{\ }H_{\mu \nu }=h_{\mu \nu }+\mathfrak{n}_{\mu \nu }M_{H}-\frac{%
\mathfrak{n}^{2}h^{2}}{2M_{H}}+O(1/M_{H}^{2})  \label{constr1}
\end{equation}%
It should be particularly emphasized that the modes collected in the $h_{\mu
\nu }$ are generally the Goldstone modes of the broken accidental $SO(7,3)$
symmetry of the constraint (\ref{const3}), thus containing the Lorentz
Goldstone modes and PGMs put together. If Lorentz symmetry is completely
broken then the pure Goldstone modes appear enough to be solely collected in
physical graviton. On the other hand, when one has a partial Lorentz
violation, some PGMs should be added.

Putting then the parameterization (\ref{constr1}) into the total Lagrangian $%
\mathcal{L}(H,A)$ given in (\ref{tl}, \ref{fp}, \ref{fh}), one comes to the
truly emergent tensor field gravity Lagrangian $\mathcal{L}(h,A)$ containing
an infinite series in powers of the $h_{\mu \nu }$ modes. For the traceless
VEV tensor $\mathfrak{n}_{\mu \nu }$ (\ref{tll}) it takes, without loss of
generality, the especially simple form \ \ \ \

\begin{eqnarray}
\mathcal{L}(h,A) &=&\frac{1}{2}\partial _{\lambda }h^{\mu \nu }\partial
^{\lambda }h_{\mu \nu }-\frac{1}{2}\partial _{\lambda }h_{tr}\partial
^{\lambda }h_{tr}-\partial _{\lambda }h^{\lambda \nu }\partial ^{\mu }h_{\mu
\nu }+\partial ^{\nu }h_{tr}\partial ^{\mu }h_{\mu \nu }+  \notag \\
&&-\frac{\mathfrak{n}^{2}}{M_{H}}h^{2}\mathfrak{n}^{\mu \lambda }\left[
\partial _{\lambda }\partial ^{\nu }h_{\mu \nu }-\frac{1}{2}\partial _{\mu
}\partial _{\lambda }h_{tr}\right] +\frac{\mathfrak{n}^{2}}{8M_{H}^{2}}%
\left( \eta ^{\mu \nu }-\frac{\mathfrak{n}^{\mu \lambda }\mathfrak{n}^{\nu
\lambda }}{\mathfrak{n}^{2}}\right) \partial _{\mu }h^{2}\partial _{\nu
}h^{2}  \notag \\
&&+\mathcal{L}(A)+\frac{M_{H}}{M_{P}}\left[ \mathfrak{n}_{\mu \nu }F^{\mu
\rho }F_{\rho }^{\nu }\right] -\frac{1}{M_{P}}h_{\mu \nu }T^{\mu \nu }-\frac{%
1}{2M_{H}M_{P}}h^{2}\left[ \mathfrak{n}_{\mu \nu }F^{\mu \rho }F_{\rho
}^{\nu }\right]  \label{gl}
\end{eqnarray}%
written in the $O(h^{2}/M_{H}^{2})$ approximation in which, besides the
conventional graviton bilinear kinetic terms, there are also three- and
four-linear interaction terms in powers of $h_{\mu \nu }$ in the Lagrangian.
Some of the notations used are collected below

\begin{equation}
h^{2}\equiv h_{\mu \nu }h^{\mu \nu }\text{ , \ \ }h_{tr}\equiv \eta ^{\mu
\nu }h_{\mu \nu }\text{ \ }  \label{n}
\end{equation}

The bilinear vector field term \ \

\begin{equation}
\frac{M_{H}}{M_{P}}\left[ \mathfrak{n}_{\mu \nu }F^{\mu \rho }F_{\rho }^{\nu
}\right]  \label{t}
\end{equation}%
in the third line in the Lagrangian (\ref{gl}) merits special notice. This
term arises from the interaction Lagrangian $\mathcal{L}_{int}$ (\ref{fh})
after application of the tracelessness condition (\ref{tll}) for the VEV
tensor $\mathfrak{n}_{\mu \nu }$. It could significantly affect the
dispersion relation for the vector field $A$ (and any other matter as well)
thus leading to an unacceptably large Lorentz violation if the SLIV scale $%
M_{H}$ were comparable with the Planck mass $M_{P}$. However, this term can
be gauged away \cite{cjt} by an appropriate redefinition of the vector field
by going to the new coordinates%
\begin{equation}
x^{\mu }\rightarrow x^{\mu }+\xi ^{\mu }.  \label{coo}
\end{equation}%
In fact, with a simple choice of the parameter function $\xi ^{\mu }(x)$
being linear in 4-coordinate

\begin{equation}
\xi ^{\mu }(x)=\frac{M_{H}}{M_{P}}\mathfrak{n}^{\mu \nu }x_{\nu }\text{ ,}
\label{df}
\end{equation}%
the term (\ref{t}) is cancelled by an analogous term stemming from the
vector field kinetic term in $\mathcal{L}(A)$ given in (\ref{fh}). On the
other hand, since the diff invariance is an approximate symmetry of the
Lagrangian $\mathcal{L}(H,A)$ we started with (\ref{tl}), this cancellation
will only be accurate up to the linear order corresponding to the tensor
field theory. Indeed, a proper extension of this theory to GR$^{1}$ with its
exact diff invariance will ultimately restore the usual dispersion relation
for the vector (and other matter) fields. We will consider all that in
significant detail in the next section.

Together with the Lagrangian one must also specify other supplementary
conditions for the tensor field $h^{\mu \nu }$(appearing eventually as
possible gauge fixing terms in the emergent tensor field gravity) in
addition to the basic emergent \ "gauge" \ condition $\mathfrak{n}_{\mu \nu
}h^{\mu \nu }=0$ given above (\ref{par}). The point is that the spin $1$
states are still left in the theory being described by some of the
components of \ the new tensor $h_{\mu \nu }.$ This is certainly
inadmissible \footnote{%
Indeed, the spin-$1$ component must be necessarily excluded in the tensor $%
h_{\mu \nu }$, since the sign of the energy for the spin-$1$ component is
always opposite to that for the spin-$2$ and spin-$0$ ones.}. Usually, the
spin $1$ states (and one of the spin $0$ states) are excluded by the
conventional Hilbert-Lorentz condition

\begin{equation}
\partial ^{\mu }h_{\mu \nu }+q\partial _{\nu }h_{tr}=0  \label{HL}
\end{equation}%
($q$ is an arbitrary constant, giving for $q=-1/2$ the standard harmonic
gauge condition). However, as we have already imposed the emergent
constraint (\ref{par}), we can not use the full Hilbert-Lorentz condition (%
\ref{HL}) eliminating four more degrees of freedom in $h_{\mu \nu }.$
Otherwise, we would have an "over-gauged" theory with a non-propagating
graviton. In fact, the simplest set of conditions which conform with the
emergent condition $\mathfrak{n}\cdot h=0$ in (\ref{par}) turns out to be

\begin{equation}
\partial ^{\rho }(\partial _{\mu }h_{\nu \rho }-\partial _{\nu }h_{\mu \rho
})=0  \label{gauge}
\end{equation}%
This set excludes only three degrees of freedom \footnote{%
The solution for a gauge function $\xi _{\mu }(x)$ satisfying the condition$%
\ $(\ref{gauge}) $\ $\ can generally be chosen as $\xi _{\mu }=$\ $\ \square
^{-1}(\partial ^{\rho }h_{\mu \rho })+\partial _{\mu }\theta $, where $%
\theta (x)$ is an arbitrary scalar function, so that only three degrees of
freedom in $h_{\mu \nu }$ are actually eliminated.} in $h_{\mu \nu }$ and,
besides, it automatically satisfies the Hilbert-Lorentz spin condition as
well. So, with the Lagrangian (\ref{gl}) and the supplementary conditions\ (%
\ref{par}) and (\ref{gauge}) lumped together, one eventually comes to a
working model for the emergent tensor field gravity \cite{cjt}. Generally,
from ten components of the symmetric two-index tensor $h_{\mu \nu }$ four
components are excluded by the supplementary conditions (\ref{par}) and (\ref%
{gauge}). For a plane gravitational wave propagating in, say, the $z$
direction another four components are also eliminated, due to the fact that
the above supplementary conditions still leave freedom in the choice of a
coordinate system, $x^{\mu }\rightarrow $ $x^{\mu }+\xi ^{\mu }(t-z/c),$
much as it takes place in standard GR. Depending on the form of the VEV
tensor $\mathfrak{n}_{\mu \nu }$, caused by SLIV, the two remaining
transverse modes of the physical graviton may consist solely of Lorentzian
Goldstone modes or of pseudo-Goldstone modes, or include both of them. This
theory, similar to the nonlinear QED \cite{nambu}, while suggesting an
emergent description for graviton, does not lead to physical Lorentz
violation \cite{cjt}.

\section{Electrogravity theory}

\subsection{Emergent photons and gravitons together}

So far we considered the vector field $A_{\mu }$ as an unconstrained
material field which the emergent gravitons interact with. Now, we propose
that the vector field also develops the VEV through the SLIV constraint (\ref%
{const}), thus generating the massless vector Goldstone modes\ associated
with a photon. We also include the complex scalar field $\varphi $ (taken to
be massless, for simplicity) as an actual matter in the theory%
\begin{equation}
\mathcal{L}(\varphi )=D_{\mu }\varphi \left( D_{\mu }\varphi \right) ^{\ast
},\text{ }D_{\mu }=\partial _{\mu }+ieA_{\mu }\text{ .}  \label{fi}
\end{equation}%
So, the total starting electrogravigy Lagrangian is, therefore, proposed to
be
\begin{equation}
\mathcal{L}_{tot}=\mathcal{L}(A)+\mathcal{L}(H)+\mathcal{L}(\varphi )+%
\mathcal{L}_{int}(H,A,\varphi )  \label{tot}
\end{equation}%
where the Lagrangians $\mathcal{L}(A)$ and $\mathcal{L}(H)$ were given above
in (\ref{tt}, \ref{fp}), while the gravity interaction part%
\begin{equation}
\mathcal{L}_{int}(H,A,\varphi )=-\frac{1}{M_{P}}H_{\mu \nu }[T^{\mu \nu
}(A)+T^{\mu \nu }(\varphi )]  \label{in}
\end{equation}%
contains the tensor field couplings with canonical energy-momentum tensors
of vector and scalar fields.

In the symmetry broken phase one goes to the pure Goldstone vector and
tensor modes, $a_{\mu }$ and $h_{\mu \nu }$, respectively. Whereas the
tensor modes $h_{\mu \nu }$ including their kinetic and interaction terms
have been thoroughly dicussed in the previos section (\ref{gl}), the vector
modes $a_{\mu }$ are not yet properly exposed. Puting the parametrization (%
\ref{gol}) into the Lagrangian (\ref{tt}) one has from the vector field
kinetic term (taken to the first order in $a^{2}/M_{A}^{2}$)
\begin{equation}
\mathcal{L}(A)\rightarrow \mathcal{L}(a)=-\frac{1}{4}f_{\mu \nu }f^{\mu \nu
}-\frac{1}{2}\delta (n\cdot a)^{2}-\frac{1}{4}\frac{n^{2}}{M_{A}}f_{\mu \nu
}(\partial ^{\mu \nu }a^{2})\text{ .}  \label{nl}
\end{equation}%
We have denoted the $a_{\mu }$ strength tensor by $f_{\mu \nu }=\partial
_{\mu }a_{\nu }-\partial _{\nu }a_{\mu }$, while $\partial ^{\mu \nu
}=n^{\mu }\partial ^{\nu }-n^{\nu }\partial ^{\mu }$ is a new SLIV oriented
differential tensor acting on the infinite series in $a^{2}$ coming from the
expansion of the effective \textquotedblleft Higgs" mode $%
(M_{A}^{2}-n^{2}a^{2})^{\frac{1}{2}}$ in (\ref{gol}), from which we have
only included the lowest order term $-n^{2}a^{2}/2M_{A}$ throughout the
Lagrangian $\mathcal{L}(a)$. We have also explicitly introduced in the
Lagrangian the\ emergent orthogonality condition $n\cdot a=0$ which can be
treated as the gauge fixing term (when taking the limit $\delta \rightarrow
\infty $). At the same time, the scalar field Lagrangian $\mathcal{L}%
(\varphi )$ in (\ref{tot}) is going now to
\begin{equation}
\mathcal{L}(\varphi )=\left\vert \left( \partial _{\mu }+iea_{\mu
}+ieM_{A}n_{\mu }-ie\frac{n^{2}}{2M_{A}}a^{2}n_{\mu }\right) \varphi
\right\vert ^{2}  \label{ls}
\end{equation}%
while tensor field interacting terms (\ref{in}) in $\mathcal{L}%
_{int}(H,A,\varphi )$ convert to%
\begin{equation}
\mathcal{L}_{int}=-\frac{1}{M_{P}}\left( h_{\mu \nu }+M_{H}\mathfrak{n}_{\mu
\nu }-\frac{\mathfrak{n}^{2}}{2M_{H}}h^{2}\mathfrak{n}_{\mu \nu }\right) %
\left[ T^{\mu \nu }\left( a_{\mu }-\frac{n^{2}}{2M_{A}}a^{2}n_{\mu }\right)
+T^{\mu \nu }(\varphi )\right]  \label{lin}
\end{equation}%
where the vector field energy-momentum tensor is now solely a function of
the Goldstone $a_{\mu }$ modes.

\subsection{Constraints and zero mode spectrum}

Before going any further, let us make some necessary comments. Note first of
all that, apart from dynamics decribed by the total Lagrangian $\mathcal{L}%
_{tot},$ the vector and tensor field constraints (\ref{const}, \ref{const3})
are also proposed to be satisfied. In principle, these constraints could be
formally obtained from the convential potential terms included in the total
Lagrangian $\mathcal{L}_{tot}$, as was discussed in section 1. The most
general potential, where the vector and tensor field couplings possess the
Lorentz and $SO(7,3)$ symmetry, respectively, must be solely a function of $%
A_{\mu }^{2}\equiv A_{\mu }A^{\mu }$\ and $H_{\mu \nu }^{2}\equiv H_{\mu \nu
}H^{\mu \nu }$. Indeed, it cannot include any contracted and intersecting
terms like as $H_{tr}$, \ $H^{\mu \nu }A_{\mu }A_{\nu }$ and others which
would immediately reduce the above symmetries to the common Lorentz one. So,
one may only write
\begin{equation}
U(A,H)=\lambda _{A}(A_{\mu }^{2}-n^{2}M_{A}^{2})^{2}+\lambda _{H}(H_{\mu \nu
}^{2}-\mathfrak{n}^{2}M_{H}^{2})^{2}+\lambda _{AH}A_{\mu }^{2}H_{\rho \nu
}^{2}  \label{u}
\end{equation}%
where $\lambda _{A,H,AH}$ stand for the coupling constants of the vector and
tensor fields, while values of $n^{2}=\pm 1$ and $\mathfrak{n}^{2}=\pm 1$
determine their possible vacuum configurations. As a consequence, an
absolute minimum of the potential (\ref{u}) might appear for the couplings
satisfying the conditions%
\begin{equation}
\lambda _{A,H}>0\text{ , \ }\lambda _{A}\lambda _{H}>\lambda _{AH}/4
\end{equation}%
However, as in the pure vector field case discussed in section 1, this
theory is generally unstable with the Hamiltonian being unbounded from below
unless the phase space is constrained just by the above nonlinear conditions
(\ref{const}, \ref{const3}). They in turn follow from the potential (\ref{u}%
) when going to the nonlinear $\sigma $-model type limit $\lambda
_{A,H}\rightarrow \infty $. In this limit, the massive Higgs mode disappears
from the theory, the Hamiltonian becomes positive, and one comes to the pure
emergent electrogravity theory considered here.

We note again that the Goldstone modes appearing in the theory are caused by
breaking of global symmetries related to the constraints (\ref{const}, \ref%
{const3}) rather than directly to Lorentz violation. Meanwhile, for the
vector field case symmetry of the constraint (\ref{const}) coincides in fact
with Lorentz symmetry whose breaking causes the Goldstone modes depending on
the vacuum orientation vector $n_{\mu }$, as can be clearly seen from an
appropriate exponential parametrization for the starting vector field
\begin{equation}
A_{\alpha }=\left[ e^{in_{\mu }\mathcal{J}^{\mu \nu }a_{\nu }/M_{A}}\right]
_{\alpha }^{\beta }n_{\beta }M_{A}  \label{np1}
\end{equation}%
where $n_{\mu }\mathcal{J}^{\mu \nu }$\ just corresponds to the broken
Lorentz generators. However, in the tensor field case, due to the higher
symmetry $SO(7,3)$\ of the constraint (\ref{const3}), there are much more
tensor zero modes than would appear from SLIV itself. In fact, they complete
the whole tensor multiplet $h_{\mu \nu }$ in the parametrization (\ref{par}%
). However, as was discussed in the previous section, only a part of them
are true Goldstone modes, others are pseudo-Goldstone ones. In the minimal
VEV configuration case, when these VEVs are developed only on the single $%
A_{\mu }$ and $H_{\mu \nu }$ components, one has several possibilities
determined by the vacuum orientations $n_{\mu }$ and $\mathfrak{n}_{\mu \nu
} $ in the equations (\ref{np1})\ and (\ref{ns}, \ref{nss}, \ref{np}),
respectively. There apppear the twelve zero modes in total, three from
Lorentz violation itself and nine from a violation of the $SO(7,3)$\
symmetry that is more than enough to have the necessary three photon modes
(two physical and one auxiliary ones) and six graviton modes (two physical
and four auxiliary ones). We list below all interesting cases classifying
them according to the corresponding $n-\mathfrak{n}$ values.

(1) For the timelike-timelike SLIV, when both $n_{0}\neq 0$ and $\mathfrak{n}%
_{00}\neq 0$, photon is determined by the space Goldstone components $a_{i}$
($i=1,2,3$) of the partially broken Lorentz symmetry $SO(1,3)\rightarrow
SO(3)$, while space-space components $h_{ij}$ needed for physical graviton
and its auxiliary components can be only provided by the pseudo-Goldstone
modes following from the timelike symmetry breaking $SO(7,3)$ $\rightarrow
SO(6,3)$ related to the tensor-field constraint (\ref{const3}).

(2) Another interesting case seems to be the timelike-spacelike SLIV, when $%
n_{0}\neq 0$ and $\mathfrak{n}_{i=j}\neq 0$ (one of the diagonal space
components of the unit tensor $\mathfrak{n}_{\mu \nu }$ is nonzero). Now,
Lorentz symmetry is broken up to the plane rotations $SO(1,3)\rightarrow
SO(2),$ so that the five true Goldstone bosons appear shared among photon
and graviton in the following way. Photon is given again by three space
components $a_{i}$, while graviton is determined by two space-space
components, $h_{12}$ and $h_{13}$ (if the VEV was developed along the
direction $\mathfrak{n}_{11}$), as directly follows from the parametrization
equations (\ref{np1})\ and (\ref{np}). Thus again one necessary component $%
h_{23}$ for physical graviton, as well as its gauge degrees of freedom,
should be provided by the proper pseudo-Goldstone modes following from the
spacelike symmetry breaking $SO(7,3)$ $\rightarrow SO(7,2)$ related to the
tensor-field constraint (\ref{const3}).

(3) For the similar timelike-spacelike SLIV case, when $n_{0}\neq 0$ and $%
\mathfrak{n}_{i\neq j}\neq 0$ (one of the nondiagonal space components of
the unit tensor $\mathfrak{n}_{\mu \nu }$ is nonzero) Lorentz symmetry
appears fully broken so that the photon has the same three space components $%
a_{i}$, while the graviton physical components are given by the tensor field
space components $h_{ij}$. This is the only case when all physical
components of both photon and graviton are provided by the true SLIV
Goldstone modes, whereas some gauge degrees of freedom for a graviton are
given by the PGM states stemming from the spacelike symmetry breaking $%
SO(7,3)$ $\rightarrow SO(7,1)$ related to the tensor field constraint (\ref%
{const3}).

(4) Using the parametrization equations (\ref{np1})\ and (\ref{np}) one can
readily consider all other possibilities as well, particularly, the
spacelike-timelike (nonzero $n_{i}$ and $\mathfrak{n}_{00}$),
spacelike-spacelike diagonal (nonzero $n_{i}$ and $\mathfrak{n}_{i=j}$) and
spacelike-spacelike nondiagonal (nonzero $n_{i}$ and $\mathfrak{n}_{i\neq j}$%
) cases. In all these cases, while photon may only contain true Goldstone
modes, some pseudo-Goldstone modes appear necessary to be collected in
graviton together with some true Goldstone modes.

\subsection{Emergent electrogravity interactions}

To proceed further, one should eliminate, first of all, the large terms of
the false Lorentz violation being proportional to the SLIV scales $M_{A}$
and $M_{H}$ in the interaction Lagrangians (\ref{ls}) and (\ref{lin}).
Arranging the phase transformation for the scalar field in the following way
\begin{equation}
\varphi \rightarrow \varphi \exp [-ieM_{A}n_{\mu }x^{\mu }]  \label{ph}
\end{equation}%
one can simply cancel that large term in the scalar field Lagrangian (\ref%
{ls}), thus coming to
\begin{equation}
\mathcal{L}(\varphi )=\left\vert \left( D_{\mu }-ie\frac{n^{2}}{2M_{A}}%
a^{2}n_{\mu }\right) \varphi \right\vert ^{2}  \label{ls1}
\end{equation}%
where the covariant derivative $D_{\mu }$ is read from now on as $D_{\mu
}=\partial _{\mu }+iea_{\mu }$. Another unphysical set of terms, like the
already discussed term (\ref{t}), may appear from the gravity interaction
Lagrangian $L_{int}$ (\ref{lin}) where the large SLIV\ entity $M_{H}%
\mathfrak{n}_{\mu \nu }$ couples to the energy-momentum tensor. They also
can be eliminated by going to the new coordinates (\ref{coo}), as was
mentioned in the previous section.

For infinitesimal translations $\xi _{\mu }(x)$ the tensor field transforms
according to (\ref{tr3}), while scalar and vector fields transform as%
\begin{equation}
\delta \varphi =\xi _{\mu }\partial ^{\mu }\varphi ,\text{ }\delta a_{\mu
}=\xi _{\lambda }\partial ^{\lambda }a_{\mu }+\partial _{\mu }\xi _{\nu
}a^{\nu }\text{ ,}  \label{st}
\end{equation}%
respectively. One can see, therefore, that the scalar field transformation
has only the translation part, while the vector one has an extra term
related to its nontrivial Lorentz structure. For the constant unit vector $%
n_{\mu }$ this transformation looks as
\begin{equation}
\delta n_{\mu }=\partial _{\mu }\xi _{\nu }n^{\nu },  \label{nt}
\end{equation}%
having no the translation part. Using all that and also expecting that the
phase parameter $\xi _{\lambda }$ is in fact linear in coordinate $x_{\mu }$
(that allows to drop out its high-derivative terms), we can easily calculate
all scalar and vector field variations, such as
\begin{equation}
\delta \left( D_{\mu }\varphi \right) =\xi _{\lambda }\partial ^{\lambda
}(D_{\mu }\varphi )+\partial _{\mu }\xi _{\lambda }D^{\lambda }\varphi ,%
\text{ }\delta f_{\mu \nu }=\xi _{\lambda }\partial ^{\lambda }f_{\mu \nu
}+\partial _{\mu }\xi ^{\lambda }f_{\lambda \nu }+\partial _{\nu }\xi
^{\lambda }f_{\mu \lambda }  \label{var}
\end{equation}%
and others. This finally leads to the total variations of the above
Lagrangians. Whereas the pure tensor field Lagrangian $\mathcal{L}(H)$ (\ref%
{fp})\ is invariant under diff transformations, $\delta \mathcal{L}(H)=0$,
the interaction Lagrangian $\mathcal{L}_{int}$ in (\ref{tot}) is only
approximately invariant being compensated (in the lowest order in the
transformation parameter $\xi _{\mu })$ by kinetic terms of all the fields
involved. However, this Lagrangian becomes increasingly invariant once our
theory is extending to GR$^{1}$.

In contrast, the vector and scalar field Lagrangians acquire some nontrivial
additions%
\begin{eqnarray}
\delta \mathcal{L}(A) &=&\xi _{\lambda }\partial _{\lambda }\mathcal{L}(A)-%
\frac{1}{2}\left( \partial _{\mu }\xi _{\lambda }+\partial _{\lambda }\xi
_{\mu }\right) \left[ f^{\mu \nu }f_{\nu }^{\lambda }+\frac{n^{2}}{M_{A}}%
\left( f_{\nu }^{\lambda }\partial ^{\mu \nu }a^{2}+\frac{1}{2}f_{\rho \nu
}\partial ^{\rho \nu }\left( a^{\mu }a^{\lambda }\right) \right) \right]
\notag \\
\text{ }\delta \mathcal{L}(\varphi ) &=&\xi _{\lambda }\partial _{\lambda }%
\mathcal{L}(\varphi )+\left( \partial _{\mu }\xi _{\nu }+\partial _{\nu }\xi
_{\mu }\right) \left[ \left( \mathfrak{D}^{\mu }\varphi \right) ^{\ast }%
\mathfrak{D}^{\nu }\varphi +\frac{a^{\mu }a^{\nu }n^{2}}{2M_{A}}n_{\lambda
}J_{\lambda }\right]  \label{ads}
\end{eqnarray}%
where $J_{\mu }$ stands for the conventional vector field source current
\begin{equation}
J_{\mu }=ie[\varphi ^{\ast }D_{\mu }\varphi -\varphi \left( D_{\mu }\varphi
\right) ^{\ast }]  \label{j}
\end{equation}%
while $\mathfrak{D}_{\nu }\varphi $ is the SLIV extended covariant
derivative for the scalar field
\begin{equation}
\mathfrak{D}_{\nu }\varphi =D_{\nu }\varphi -ie\frac{n^{2}}{2M_{A}}%
a^{2}n_{\nu }\varphi \text{ \ \ }  \label{not}
\end{equation}%
The first terms in the variations (\ref{ads}) \ are unessential since they
simply show that these Lagrangians transform, as usual, like as scalar
densities under diff transformations.

Combining these variations with $\mathcal{L}_{int}$ (\ref{lin}) in the total
Lagrangian (\ref{tot}) one finds after simple, though long, calculations
that the largest Lorentz violating terms in it
\begin{equation}
-\left( \frac{M_{H}}{M_{P}}\mathfrak{n}_{\mu \nu }-\frac{\partial _{\mu }\xi
_{\lambda }+\partial _{\lambda }\xi _{\mu }}{2}\right) \left[ -f^{\mu \nu
}f_{\nu }^{\lambda }-\frac{n^{2}}{M_{A}}f_{\lambda }^{\nu }\partial ^{\mu
\lambda }a^{2}+2\mathfrak{D}^{\nu }\varphi \left( \mathfrak{D}^{\mu }\varphi
\right) ^{\ast }\right]  \label{l}
\end{equation}%
will immediately cancel if the transformation parameter is chosen exactly as
is given in (\ref{df}) in the previous section. So, with this choice we
finally have for the modified interaction Lagrangian

\begin{equation}
\mathcal{L}_{int}^{\prime }(h,a,\varphi )=-\frac{1}{M_{P}}h_{\mu \nu }T^{\mu
\nu }(a,\varphi )+\frac{1}{M_{P}M_{A}}\mathcal{L}_{1}+\frac{1}{M_{P}M_{H}}%
\mathcal{L}_{2}+\frac{M_{H}}{M_{P}M_{A}}\mathcal{L}_{3}  \label{intmod}
\end{equation}%
where
\begin{eqnarray}
\mathcal{L}_{1} &=&n^{2}h_{\mu \nu }\left[ f_{\lambda }^{\nu }\partial ^{\mu
\lambda }a^{2}-n^{\mu }J^{\nu }+\eta ^{\mu \nu }\left( -\frac{1}{4}%
f_{\lambda \rho }\partial ^{\lambda \rho }a^{2}+n^{\lambda }J_{\lambda
}\right) \right]  \notag \\
\mathcal{L}_{2} &=&\frac{1}{2}\mathfrak{n}^{2}h^{2}\mathfrak{n}_{\mu \nu }%
\left[ -f^{\mu \lambda }f_{\lambda }^{\nu }+2D^{\nu }\varphi \left( D^{\mu
}\varphi \right) ^{\ast }\right]  \notag \\
\mathcal{L}_{3} &=&n^{2}\mathfrak{n}_{\mu \lambda }\left[ \frac{1}{2}f_{\rho
\nu }\partial ^{\rho \nu }\left( a^{\mu }a^{\lambda }\right) -(a^{\mu
}a^{\lambda })n^{\nu }J_{\nu }\right]  \label{3}
\end{eqnarray}%
Thereby, apart from a conventional gravity interaction part given by the
first term in (\ref{intmod}), there are Lorentz violating couplings in $%
\mathcal{L}_{1,2,3}$ being properly suppressed by corresponding mass scales.
Note that the coupling presented in $\mathcal{L}_{3}$ between the vector and
scalar fields is solely induced by the tensor field SLIV. Remarkably, this
coupling may be in principle of the order of a normal gravity coupling or
even stronger, if \ $M_{H}>M_{A}$. However, appropriately simplifying this
coupling (and using also a full derivative identity) one comes to%
\begin{equation}
\mathcal{L}_{3}\sim n^{2}\left( \mathfrak{n}_{\mu \lambda }a^{\mu
}a^{\lambda }\right) n^{\rho }\left[ \partial ^{\nu }f_{\nu \rho }-J_{\rho }%
\right]
\end{equation}%
that after applying of the vector \ field equation of motion turns it into
zero. We consider it in more detail in the next section where we calculate
some tree level processes.

\section{The lowest order SLIV processes}

\subsection{Preamble}

The emergent gravity Lagrangian in (\ref{gl}) taken alone or considered
together with the material vector and scalar fields presents in fact highly
nonlinear theory which contains lots of Lorentz and $CPT$ violating
couplings. Nevertheless, as it was shown in \cite{cjt} in the lowest order
calculations, they all are cancelled and do not manifest themselves in
physical processes. This may mean that the length-fixing constraints ( \ref%
{const3})\ put on the tensor fields appear as the gauge fixing conditions
rather than a source of an actual Lorentz violation.

However, as was mentioned in section 1, one can not be sure that these
calculations, as well as the similar calculations in the Nambu model itself,
fully confirm gauge invariance of the emergent theory considered. Indeed,
whether the constraint (\ref{const}) in QED amounts in general to a special
gauge choice for a vector field $A_{\mu }(x)$ is an open question unless the
corresponding gauge function $\omega (x)$ satisfying the constraint
condition
\begin{equation}
\lbrack A_{\mu }(x)+\partial _{\mu }\omega (x)]^{2}=n^{2}M_{A}^{2}  \label{g}
\end{equation}%
is explicitly constructed for an arbitrary $A_{\mu }(x)$. An original Nambu
argument \cite{nambu} was related to an observation that for the positive $%
n^{2}$ the constraint equation (\ref{g}) is mathematically equivalent to a
classical Hamilton-Jacobi equation for a massive charged particle
\begin{equation}
\lbrack \partial _{\mu }S(x)+eA_{\mu }(x)]^{2}=m^{2}  \label{hj}
\end{equation}%
where $S(x)$ is an action of a system, while $e$ and $m$ stand for the
particle charge and mass, respectively. Comparison of the equations (\ref{g}%
) and (\ref{hj}) shows the correspondence $\omega (x)=S(x)/e$ and $%
n^{2}M_{A}^{2}=m^{2}/e^{2}$. Thus, the constraint equation (\ref{g}) should
have a solution inasmuch there is a solution to the classical problem
described by the equation (\ref{hj}). This conclusion was actually confirmed
by Nambu for the timelike SLIV ($n^{2}=+1$) in the lowest order calculation
of the physical processes in \cite{nambu} and then was extended to the
one-loop approximation and for both\ the timelike ($n^{2}>0$) and space-like
($n^{2}<0$) Lorentz violation in \cite{az}. Thus, status of the constraint (%
\ref{const}) as a special gauge choice in QED is only partially approved by
some low order calculations rather than has a serious theoretical reason.

The same may be said about the emergent tensor field gravity. Its diff gauge
invariance could only be fully approved if the corresponding gauge function $%
\xi _{\mu }(x)$ satisfying the constraint condition (\ref{const3})
\begin{equation}
\lbrack H_{\mu \nu }(x)+\partial _{\mu }\xi _{\nu }(x)+\partial _{\nu }\xi
_{\mu }]^{2}=\mathfrak{n}^{2}M_{H}^{2}  \label{gg}
\end{equation}%
is explicitly constructed for an arbitrary $H_{\mu \nu }(x)$. However, in
contrast to the above nonlinear QED case where at least some heuristic
argument could be applied, one cannot be sure that there exist a solution to
the equation (\ref{gg}) in a general case. So, the only way to answer this
question is to explicitly check it in physical processes that in the lowest
approximation has been done in \cite{cjt}. Again, though the result appears
positive, one cannot be sure that this will work in all orders.

The present electrogravity theory, in contrast to the pure QED and tensor
field gravity theories, contains both photon and graviton as the emergent
gauge fields. This adds new variety of \ Lorentz and $CPT$ violating
couplings (\ref{intmod}) being expressed in terms of tensor and vector
Goldstone modes. In general, one cannot be sure that, even though both the
emergent QED and tensor field gravity taken separately preserve Lorentz
invairance (in the low order processes), the combined electrogravity theory
does not lead to physical lorentz violation as well. However, as shows our
calculations given below,\ just this appears to be the case. All Lorentz
violation effects turn out again to be strictly cancelled among themselves
at least in the lowest order SLIV processes in the electrogravity theory.
Thus, similar to emergent vector field theories, both Abelian \cite%
{nambu,az,kep} and non-Abelian \cite{jej}, as well as in the pure tensor
field gravity \cite{cjt}, such a cancellation may only means that at least
in the lowest approximation the SLIV constraints (\ref{const}, \ref{const3})
amount to a special gauge choices in an otherwise diff and Lorentz invariant
emergent electrogravity theory presented here.

We will consider the lowest order SLIV processes, once the corresponding
Feynman rules are properly established. For simplicity, both in the above
Lagrangians and forthcoming calculations, we continue using the traceless of
the VEV tensor $\mathfrak{n}_{\mu \nu }$ (\ref{tll}), while our results
remain true for any type of vacuum configuration caused by SLIV.

\subsection{Feynman rules}

Though the Feynman rules and processes related to the nonlinear QED, as well
as with emergent gravity with the matter scalar fields, are thoroughly
discussed in our previous works \cite{cjt,kep}, there are many new Lorentz
and $CPT$ breaking interactions in the total interaction Lagrangian (\ref%
{intmod}). We present below some basic Feynman rules which are needed for
calculations of different SLIV processes just appearing in the emergent
electrogravity.

(\textbf{i}) The first and most important is the graviton propagator which
only conforms with the emergent gravity Lagrangian (\ref{gl}) and the gauge
conditions\ (\ref{par}) and (\ref{gauge})

\begin{eqnarray}
-iD_{\mu \nu \alpha \beta }\left( k\right) &=&\frac{1}{2k^{2}}\left( \eta
_{\beta \mu }\eta _{\alpha \nu }+\eta _{\beta \nu }\eta _{\alpha \mu }-\eta
_{\alpha \beta }\eta _{\mu \nu }\right)  \notag \\
&&-\frac{1}{2k^{4}}\left( \eta _{\beta \nu }k_{\alpha }k_{\mu }+\eta
_{\alpha \nu }k_{\beta }k_{\mu }+\eta _{\beta \mu }k_{\alpha }k_{\nu }+\eta
_{\alpha \mu }k_{\beta }k_{\nu }\right)  \label{prop} \\
&&-\frac{1}{k^{2}(\mathfrak{n}kk)}\left( k_{\alpha }k_{\beta }\mathfrak{n}%
_{\mu \nu }+k_{\nu }k_{\mu }\mathfrak{n}_{\alpha \beta }\right) +\frac{1}{%
k^{2}(\mathfrak{n}kk)^{2}}\left[ \mathfrak{n}^{2}-\frac{2}{k^{2}}(k\mathfrak{%
nn}k)\right] k_{\mu }k_{\nu }k_{\alpha }k_{\beta }  \notag \\
&&+\frac{1}{k^{4}(\mathfrak{n}kk)}\left( \mathfrak{n}_{\mu \rho }k^{\rho
}k_{\nu }k_{\alpha }k_{\beta }+\mathfrak{n}_{\nu \rho }k^{\rho }k_{\mu
}k_{\alpha }k_{\beta }+\mathfrak{n}_{\alpha \rho }k^{\rho }k_{\nu }k_{\mu
}k_{\beta }+\mathfrak{n}_{\beta \rho }k^{\rho }k_{\nu }k_{\alpha }k_{\mu
}\right)  \notag
\end{eqnarray}%
(where $(\mathfrak{n}kk)\equiv \mathfrak{n}_{\mu \nu }k^{\mu }k^{\nu }$ and $%
(k\mathfrak{nn}k)\equiv k^{\mu }\mathfrak{n}_{\mu \nu }\mathfrak{n}^{\nu
\lambda }k_{\lambda })$. It automatically satisfies the orthogonality
condition $\mathfrak{n}^{\mu \nu }D_{\mu \nu \alpha \beta }\left( k\right)
=0 $ and on-shell transversality $k^{\mu }k^{\nu }D_{\mu \nu \alpha \beta
}(k,k^{2}=0)=0.$ $\ $This is consistent with the corresponding polarization
tensor $\epsilon _{\mu \nu }(k,k^{2}=0)$ of the free tensor fields, being
symmetric$,$ transverse ($k^{\mu }\epsilon _{\mu \nu }=0),$traceless ($\eta
^{\mu \nu }\epsilon _{\mu \nu }(k)=0$) and also orthogonal to the vacuum
direction, $\mathfrak{n}^{\mu \nu }\epsilon _{\mu \nu }(k)=0$. As one can
see, only standard terms given by the first bracket in (\ref{prop})
contribute when the propagator is sandwiched between the conserved
energy-momentum tensors of matter fields, and the result is always Lorentz
invariant.

We will also need the photon propagator
\begin{equation}
ik^{2}D_{\mu \nu }=\eta _{\mu \nu }-\frac{n_{\mu }k_{\nu }+n_{\nu }k_{\mu }}{%
n\cdot k}+\frac{n^{2}}{\left( n\cdot k\right) ^{2}}k_{\mu }k_{\nu }
\label{propa}
\end{equation}%
which in accordance with vector field Lagrangian (\ref{nl}) possesses the
following properties: $n^{\mu }D_{\mu \nu }=0$ and $k^{\mu }D_{\mu \nu
}(k^{2}=0)=0$.

(\textbf{ii}) Next is the 3-graviton vertex $hhh$, again from the Lagrangian
(\ref{gl}), with graviton polarization tensors (and 4-momenta) given by $%
\epsilon ^{\alpha \alpha ^{\prime }}(k_{1}),$ $\epsilon ^{\beta \beta
^{\prime }}(k_{2})$ and $\epsilon ^{\gamma \gamma ^{\prime }}(k_{3})$

\begin{eqnarray}
\Gamma _{3h}^{\alpha \alpha ^{\prime }\beta \beta ^{\prime }\gamma \gamma
^{\prime }} &=&\frac{i}{2M_{H}}[\left( \eta ^{\beta \gamma }\eta ^{\beta
^{\prime }\gamma ^{\prime }}+\eta ^{\beta \gamma ^{\prime }}\eta ^{\beta
^{\prime }\gamma }\right) P^{\alpha \alpha ^{\prime }}(k_{1})  \notag \\
&&+\left( \eta ^{\alpha \gamma }\eta ^{\alpha ^{\prime }\gamma ^{\prime
}}+\eta ^{\alpha \gamma ^{\prime }}\eta ^{\alpha ^{\prime }\gamma }\right)
P^{\beta \beta ^{\prime }}(k_{2})  \label{3h} \\
&&+\left( \eta ^{\beta \alpha }\eta ^{\beta ^{\prime }\alpha ^{\prime
}}+\eta ^{\beta \alpha ^{\prime }}\eta ^{\beta ^{\prime }\alpha }\right)
P^{\gamma \gamma ^{\prime }}(k_{3})  \notag
\end{eqnarray}%
where the momentum tensor $P^{\mu \nu }(k)$ is
\begin{equation}
P^{\mu \nu }(k)=\mathfrak{n}^{\nu \rho }k_{\rho }k^{\mu }+\mathfrak{n}^{\mu
\rho }k_{\rho }k^{\nu }-\eta ^{\mu \nu }(\mathfrak{n}kk)\text{ }  \label{p}
\end{equation}%
\ Note that all 4-momenta at the vertices are taken ingoing throughout.

\textbf{(iii) } Then, the contact tensor-tensor-vector-vector interaction
coupling $hhaa$ coming from the Lagrangian $\mathcal{L}_{2}$ in (\ref{3}).
However, \ it would be useful to give first the standard
tensor-vector-vector vertex $haa$ with tensor and vector field
polarizations, $\epsilon ^{\alpha \alpha ^{\prime }}$ and $\xi ^{\mu ,\nu }$%
, respectively,
\begin{equation}
\Gamma _{st}^{\alpha \alpha ^{\prime }\mu \nu }=-\frac{i}{M_{P}}T^{\alpha
\alpha ^{\prime }\mu \nu }(a_{\lambda })  \label{h2a}
\end{equation}%
where $T^{\alpha \alpha ^{\prime }\mu \nu }(a_{\lambda })$ stands for the
conserved energy-momentum tensor of Goldstone vector field $a_{\lambda }$

\begin{eqnarray}
T^{\alpha \alpha ^{\prime }\mu \nu } &=&\frac{1}{2}\left( \eta ^{\alpha \mu
^{\prime }}\eta ^{\alpha ^{\prime }\nu ^{\prime }}+\eta ^{\alpha \nu
^{\prime }}\eta ^{\alpha ^{\prime }\mu ^{\prime }}\right) [\left( k_{2\mu
^{\prime }}\eta _{\nu }^{\lambda }-k_{2}^{\lambda }\eta _{\mu ^{\prime }\nu
}\right) \left( k_{1\nu ^{\prime }}\eta _{\lambda \mu }-k_{1\lambda }\eta
_{\mu \nu ^{\prime }}\right)  \notag \\
&&+\left( k_{2\nu ^{\prime }}\eta _{\nu }^{\lambda }-k_{2}^{\lambda }\eta
_{\nu ^{\prime }\nu }\right) \left( k_{1\mu ^{\prime }}\eta _{\lambda \mu
}-k_{1\lambda }\eta _{\mu \mu ^{\prime }}\right) ]  \label{c}
\end{eqnarray}%
being properly conserved
\begin{equation}
(k_{1}-k_{2})_{\alpha }T^{\alpha \alpha ^{\prime }\mu \nu }\xi _{\mu
}(k_{1})\xi _{\nu }(k_{2})=0\text{ .}  \label{cc}
\end{equation}%
One can see that this tensor is symmetric both in the ($\alpha $, $\alpha
^{\prime }$) and ($\mu $, $\nu $) indices, though these pairs are not
interchangeable. Using all that we are ready now to give the contact
tensor-tensor-vector-vector interaction vertex%
\begin{equation}
\Gamma _{2h2a}^{\beta \beta ^{\prime }\gamma \gamma ^{\prime }\mu \nu }=%
\frac{i\mathfrak{n}^{2}}{2M_{P}M_{H}}\left( \eta ^{\beta \gamma }\eta
^{\beta ^{\prime }\gamma ^{\prime }}+\eta ^{\beta \gamma ^{\prime }}\eta
^{\beta ^{\prime }\gamma }\right) \mathfrak{n}_{\alpha \alpha ^{\prime
}}T^{\alpha \alpha ^{\prime }\mu \nu }(a_{\lambda })  \label{2h2a}
\end{equation}%
with the corresponding tensor field ($\beta \beta ^{\prime }$ and $\gamma
\gamma ^{\prime }$) and vector field ($\mu $, $\nu $) polarization indices.

(\textbf{iv}) We have also to derive the 4-linear tensor-vector interaction
vertex $haaa$\ coming from the Lagrangian $\mathcal{L}_{1}$ in (\ref{3}).
Note that the last term in it which is proportional to $h_{tr}$ will not
contribute in the processes with graviton on external lines, since its
polarization tensor is traceless. For the other terms one has the vertex
\begin{eqnarray}
\Gamma _{h3a}^{\alpha \alpha ^{\prime }\mu \nu \lambda } &=&\frac{in^{2}}{%
M_{P}M_{A}}\left( \eta ^{\alpha \mu ^{\prime }}\eta ^{\alpha ^{\prime }\nu
^{\prime }}+\eta ^{\alpha \nu ^{\prime }}\eta ^{\alpha ^{\prime }\mu
^{\prime }}\right)  \notag \\
&&[\eta _{\nu \lambda }\left( \left( n\cdot k_{1}\right) \left(
p+k_{1}\right) _{\mu ^{\prime }}\eta _{\nu ^{\prime }\mu }+n_{\mu ^{\prime
}}\left( p+k_{1}\right) ^{\rho }\left( k_{1\nu ^{\prime }}\eta _{\rho \mu
}-k_{1\rho }\eta _{\nu ^{\prime }\mu }\right) \right)  \notag \\
&&+\eta _{\mu \lambda }\left( \left( n\cdot k_{2}\right) \left(
p+k_{2}\right) _{\mu ^{\prime }}\eta _{\nu ^{\prime }\nu }+n_{\mu ^{\prime
}}\left( p+k_{2}\right) ^{\rho }\left( k_{2\nu ^{\prime }}\eta _{\rho \nu
}-k_{2\rho }\eta _{\nu ^{\prime }\nu }\right) \right)  \notag \\
&&+\eta _{\nu \mu }\left( \left( n\cdot k_{3}\right) \left( p+k_{3}\right)
_{\mu ^{\prime }}\eta _{\nu ^{\prime }\lambda }+n_{\mu ^{\prime }}\left(
p+k_{3}\right) ^{\rho }\left( k_{3\nu ^{\prime }}\eta _{\rho \lambda
}-k_{3\rho }\eta _{\nu ^{\prime }\lambda }\right) \right) ]  \label{h3a}
\end{eqnarray}%
where polarization $\epsilon ^{\alpha \alpha ^{\prime }}(p)$ stands
interacting tensor field, while polarizations $\xi _{1\mu }(k_{1})$, $\xi
_{2\nu }(k_{2})$, $\xi _{3\lambda }(k_{3})$ for interacting vector fields.

(\textbf{v}) For the three-vector Goldstone mode interaction $aaa$ we have
the known vertex \cite{kep} following for the pure vector field Lagrangian (%
\ref{nl})
\begin{equation}
\Gamma _{3a}^{\mu \nu \lambda }=-i\frac{n^{2}}{M_{A}}\left[ (n\cdot
k_{1})\eta _{\nu \lambda }k_{1\mu }+(n\cdot k_{2})\eta _{\mu \lambda
}k_{2\nu }+(n\cdot k_{3})\eta _{\mu \nu }k_{3\lambda }\right]  \label{3a}
\end{equation}%
and the new one coming from the Lagrangian $\mathcal{L}_{3}$ in (\ref{3})

\begin{equation}
\Gamma _{3a}^{\prime \mu \nu \lambda }=i\frac{n^{2}M_{H}}{M_{P}M_{A}}\left[
(n\cdot k_{1})\mathfrak{n}_{\nu \lambda }k_{1\mu }+(n\cdot k_{2})\mathfrak{n}%
_{\mu \lambda }k_{2\nu }+(n\cdot k_{3})\mathfrak{n}_{\mu \nu }k_{3\lambda }%
\right]  \label{3a'}
\end{equation}

(\textbf{vi}) And finally, let us give also the vector-scalar-scalar
interaction $a\varphi \varphi ^{\ast }$ stemming from the same Lagrangian $%
\mathcal{L}_{3}$%
\begin{equation}
\Gamma _{a2\varphi }^{\prime \mu \lambda }=i\frac{en^{2}M_{H}}{M_{P}M_{A}}%
l_{\mu \lambda }n^{\nu }J_{\nu }  \label{a2s'}
\end{equation}%
where $J_{\nu }$ is the conserved scalar field current discussed in the
previous section.

These are rules that are actually needed to calculate the lowest order SLIV\
processes mentioned above. Note also that some of these processes could in
principle appear in the pure nonlinear QED \cite{kep} or in the nonlinear
tensor field gravity \cite{cjt} where, as we know, all the physical Lorentz
violation effects are eventually vanished. Therefore, we consider the SLIV
contributions which only appear in the combined nonlinear vector-tensor
electrogravity theory presented here.

\subsection{Elastic photon-graviton scattering}

This SLIV\ part of this process, $\gamma +g\rightarrow \gamma +g$, may only
go in the order of \ $1/M_{P}M_{H}$ due to the emergent nature of graviton.
There are in fact two matrix elements: the first one is related to the
contact diagram with the $hhaa$ vertex (\ref{2h2a})
\begin{equation}
\mathcal{M}_{con}=\frac{i\mathfrak{n}^{2}}{M_{P}M_{H}}\left( \epsilon
_{1}\cdot \epsilon _{2}\right) \mathfrak{n}_{\alpha \alpha ^{\prime
}}T^{\alpha \alpha ^{\prime }\mu \nu }\xi _{3\mu }\xi _{4\nu }\text{,}
\label{m01}
\end{equation}%
while the second one to the pole diagram with the longitudinal graviton
exchange between the Lorentz violating $h^{3}$ (\ref{3h}) and the standard $%
haa$ (\ref{h2a}) vertices%
\begin{equation}
\mathcal{M}_{pole}=\epsilon _{1}^{\gamma \gamma ^{\prime }}\epsilon
_{2}^{\beta \beta ^{\prime }}\Gamma _{3h}^{\alpha \alpha ^{\prime }\beta
\beta ^{\prime }\gamma \gamma ^{\prime }}D_{\alpha \alpha ^{\prime }\lambda
\rho }(q)\Gamma _{st}^{\lambda \rho \mu \nu }\xi _{3\mu }\xi _{4\nu }
\label{m02}
\end{equation}%
with the graviton and photon polarizations, $\epsilon _{1,2}$ and $\xi
_{3,4} $, respectively ($q$ is a momentum of the propagating graviton). \

Note now that all the terms in the propagator $D_{\alpha \alpha ^{\prime
}\lambda \rho }$ which are proportional to the propagating momentum will
turn the energy-momentum tensor to zero, thus there are left only a few
terms in the pole matrix element $\mathcal{M}_{2}$. Using also that in the
vertex $\Gamma _{3h}$ survives one term only when the transversality and
tracelessness of a graviton is used we come to%
\begin{eqnarray}
\mathcal{M}_{pole} &=&\frac{i\mathfrak{n}^{2}}{M_{H}}\left( \epsilon
_{1}\cdot \epsilon _{2}\right) \left[ 2\mathfrak{n}^{\alpha \rho }q_{\rho
}q^{\alpha ^{\prime }}-\eta ^{\alpha \alpha ^{\prime }}(\mathfrak{n}qq)\text{
}\right] \frac{i}{q^{2}}\cdot  \notag \\
&&\left[ \frac{\eta _{\alpha \lambda }\eta _{\alpha ^{\prime }\rho }+\eta
_{\alpha ^{\prime }\lambda }\eta _{\alpha \rho }-\eta _{\alpha \alpha
^{\prime }}\eta _{\lambda \rho }}{2}-\frac{q_{\alpha ^{\prime }}q_{\alpha }%
\mathfrak{n}_{\lambda \rho }}{\left( \mathfrak{n}qq\right) }\right] \frac{-i%
}{M_{P}}T^{\lambda \rho \mu \nu }\xi _{3\mu }\xi _{4\nu }
\end{eqnarray}%
that after evident simplifications is exactly cancelled with the contact
matrix element $\mathcal{M}_{1}$ given above in (\ref{m01})%
\begin{equation}
\mathcal{M}_{tot}=\mathcal{M}_{con}+\mathcal{M}_{pole}=0\text{ .}
\end{equation}%
Thereby, physical Lorentz invariance is left intact in the emergent
graviton-photon scattering.

\subsection{Photon-graviton conversion}

This SLIV process $\gamma +g\rightarrow \gamma +\gamma $ appears in the
order of $1/M_{A}M_{P}$ (now, due to the emergent nature of photon). Again,
this process in the tree approximation is basically related to the interplay
between the contact and pole diagrams.

The contact $haaa$ diagram being determined by the interaction vertex (\ref%
{h3a}) has a matrix element
\begin{equation}
\mathcal{M}_{con}=\epsilon _{\alpha \alpha ^{\prime }}(p)\Gamma
_{h3a}^{\alpha \alpha ^{\prime }\mu \nu \lambda }\xi _{1\mu }(k_{1})\xi
_{2\nu }(k_{2})\xi _{3\lambda }(k_{3})  \label{con}
\end{equation}%
where the polarization $\epsilon _{\alpha \alpha ^{\prime }}$ belongs to the
graviton, while polarizations $\xi _{1\mu }(k_{1})$, $\xi _{2\nu }(k_{2})$, $%
\xi _{3\lambda }(k_{3})$ to the photons ($p$ and $k_{1}$ are incoming and $%
k_{2}$ and $k_{3}$ outgoing momenta).

In turn, the pole diagrams with the longitudinal photon exchange between the
Lorentz violating $a^{3}$ (\ref{3a}) and the standard $haa$ (\ref{h2a})
vertices consist in fact of three diagrams differing from each other by the
interchangeable external photon legs. Their total matrix element is%
\begin{equation}
\mathcal{M}_{pole}=\epsilon _{\alpha \alpha ^{\prime }}\Gamma _{st}^{\alpha
\alpha ^{\prime }\mu \nu }D_{\nu \lambda }(q)\Gamma _{3a}^{\lambda \rho
\sigma }(\xi _{1\mu }\xi _{2\rho }\xi _{3\sigma }+\xi _{2\mu }\xi _{1\rho
}\xi _{3\sigma }+\xi _{3\mu }\xi _{1\rho }\xi _{2\sigma })  \label{pole}
\end{equation}%
were $q$ is the propagating momentum, while momenta of the raviton and
photons, $p$ and $k_{1,2,3}$, are meant in the polarizations $\epsilon
_{\alpha \alpha ^{\prime }}$ and $\xi _{1,2,3}$.

Using again the orthogonality properties and mass shell conditions for
polarizations of the photons and graviton one can split the contact
amplitude (\ref{con}) into three terms which exactly cancel the
corresponding terms in the pole amplitude (\ref{pole}). So, we will not have
any physical Lorentz violation in this process as well\footnote{%
Note that together with the pure QED $a^{3}$ vertex (\ref{3a}) we could also
use the new $a^{3}$ vertex (\ref{3a'}) in the above pole diagrams. This
would give some new contribution into this process with the lesser order $%
M_{H}/M_{A}M_{P}^{2}$. One may expect, however, that such a contribution
will be cancelled by the corresponding contact term appearing in the same
order when going to GR (see the footnote$^{1}$).}.

\subsection{Elastic photon-scalar scattering}

One can also consider a new type of the vector field scattering process on
the charged scalar, $\gamma +s\longrightarrow \gamma +s$, appearing in the $%
eM_{H}/M_{P}M_{A}$ order due to an emergent nature of both photon and
graviton\footnote{%
Note that a similar SLIV process can independently appear in the pure
nonlinear scalar QED including the Lagrangians (\ref{nl}) and (\ref{ls}).
However, it was shown \cite{kep} that the corresponding Lorentz violation
terms are strictly cancelled in this scattering process.}. Again, there are
contact and pole diagrams for this process which cancel each other. The
contact diagram corresponds to the vertex $a\varphi \varphi ^{\ast }$ (\ref%
{a2s'}) appearing from the Lagrangian $\mathcal{L}_{3}$ in (\ref{3}) and
leads to the matrix element%
\begin{equation}
\mathcal{M}_{con}=i\frac{en^{2}M_{H}}{M_{P}M_{A}}\left( \mathfrak{n}_{\mu
\lambda }\xi _{1\mu }\xi _{2\lambda }\right) n_{\nu }J_{\nu }  \label{c1}
\end{equation}%
Meanwhile, for the pole diagram with the longitudinal photon exchange
between the Lorentz violating $aaa$ vertex (\ref{3a'}) and the standard
scalar field current (\ref{j}) in $\mathcal{L}_{3}$ one has using mass shell
properties of the vector field polarization
\begin{equation}
\mathcal{M}_{pole}=i\frac{n^{2}M_{H}}{M_{P}M_{A}}\left( \mathfrak{n}_{\mu
\lambda }\xi _{1\mu }\xi _{2\lambda }\right) (n\cdot q)q_{\nu }D_{\nu \rho
}(q)\left( ieJ_{\rho }\right) \text{ .}  \label{p1}
\end{equation}%
This amplitude, when applying an explicit form of propagator and the current
conservation $q_{\rho }J_{\rho }=0$, is exactly cancelled with the contact
one (\ref{c1}). Therefore, we show one time more that there is no real
physical SLIV effect in the theory considered.

\subsection{Other processes}

Many other tree level Lorentz violating processes related to gravitons and
vector fields (interacting with each other and the matter scalar field in
the theory) appear in higher orders in the basic SLIV parameters $1/M_{H}$
and $1/M_{A}$, by iteration of couplings presented in our basic Lagrangians (%
\ref{gl}, (\ref{intmod})) or from a further expansions of the effective
vector and tensor field Higgs modes (\ref{constr}, \ref{constr1}) inserted
into the starting total Lagrangian (\ref{tot}). Again, their amplitudes are
essentially determined by an interrelation between the longitudinal graviton
and photon exchange diagrams and the corresponding contact interaction
diagrams, which appear to cancel each other, thus eliminating physical
Lorentz violation in the theory.

Most likely, the same conclusion could be expected for SLIV\ loop
contributions as well. Actually, as in the massless QED case considered
earlier \cite{az}, the corresponding one-loop matrix elements in our
emergent electrogravity theory could either vanish by themselves or amount
to the differences between pairs of similar integrals whose integration
variables are shifted relative to each other by some constants (being in
general arbitrary functions of the external four-momenta of the particles
involved) which, in the framework of dimensional regularization, could lead
to their total cancellation.

So, the emergent electrogravity theory considered here is likely to
eventually possess physical Lorentz invariance provided that the underlying
gauge and diff invariance in the theory remains unbroken.

\section{Conclusion}

We have developed an emergent electrogravity theory consisting of the
ordinary QED and the tensor field gravity model (which mimics the linearized
general relativity in Minkowski spacetime) where both photons and gravitons
emerge as states solely consisting of massless Goldstone and
pseudo-Goldstone modes. This \ appears due to spontaneous violation of
Lorentz symmetry incorporated into global symmetries of the length-fixing
constraints put on the starting vector and tensor fields, $A_{\mu }^{2}=\pm
M_{A}^{2}$ and $H_{\mu \nu }^{2}=\pm M_{H}^{2}$ ($M_{A}$ and $M_{H}$ are the
proposed symmetry breaking scales). While for a vector field case the
symmetry of the constraint coincides with Lorentz symmetry $SO(1,3)$ of the
electrogravity Lagrangian, the tensor field constraint itself possesses much
higher global symmetry $SO(7,3)$, whose spontaneous violation provides a
sufficient number of zero modes collected in a graviton. Accordingly, while
photon may only contain true Goldstone modes, graviton appears at least
partially composed from pseudo-Goldstone modes rather than from pure
Goldstone ones. Thereby, the SLIV pattern related to breaking of the
constraint symmetries, due to which the true vacuum in the theory is chosen,
induces a variety of zero modes shared among photon and graviton.

This theory looks essentially nonlinear and contains a variety of \ Lorentz
and $CPT$ violating couplings, when expressed in terms of the pure tensor
Goldstone modes. Nonetheless, all the SLIV effects turn out to be strictly
cancelled in the lowest order processes considered. This can be taken as an
indication that in the electrogravity theory physical Lorentz invariance is
preserved in this approximation. Thereby, the length-fixing constraints (\ref%
{const}, \ref{const3})\ put on the vector and tensor fields appear as the
gauge fixing conditions rather than sources of the actual Lorentz violation
in the gauge and diff invariant Lagrangian (\ref{tot}) we started with. In
fact, some Lorentz violation through deformed dispersion relations for the
material fields involved would appear in the interaction sector (\ref{lin}),
which only possesses an approximate diff invariance. However, a proper
extension of the tensor field theory to GR, with its exact diff invariance,
ultimately restores the normal dispersion relations and, therefore, the SLIV
effects are cancelled at least in the lowest order considered. If this
cancellation were to work in all orders, one could propose that emergent
theories, like as the electrogravity theory, are not differed from
conventional gauge theories. Accordingly, spontaneous Lorentz violation
caused by the vector and tensor field constraints (\ref{const}, \ref{const3}%
) appear hidden in gauge degrees of freedom, and only results in a
noncovariant gauge choice in an otherwise gauge invariant emergent
electrogravity theory.

From this standpoint, the only way for physical Lorentz violation to occur
would be if the above gauge invariance were slightly broken at the Planck
scale order distances that could be presumably caused by quantum gravity.
This is in fact a place where the emergent vector and tensor field theories
may drastically differ from conventional QED, Yang-Mills and GR theories
where gauge symmetry breaking could hardly induce physical Lorentz
violation. In contrast, in emergent electrogravity such breaking could
readily lead to many violation effects including deformed dispersion
relations for all matter fields involved. Another basic distinction of
emergent theories with non-exact gauge invariance is a possible origin of a
mass for graviton and other gauge fields (namely, for the non-Abelian ones,
see \cite{jej}), if they, in contrast to photon, are partially composed from
pseudo-Goldstone modes rather than from pure Goldstone ones. Indeed, these
PGMs are no longer protected by gauge invariance and may properly acquire
tiny masses, which still do not contradict experiment. This may lead to a
massive gravity theory where the graviton mass emerges dynamically, thus
avoiding the notorious discontinuity problem \cite{zvv}.

So, while emergent theories with an exact local invariance are physically
indistinguishable from conventional gauge theories, there are some principal
distinctions when this local symmetry is slightly broken which could
eventually allow us to differentiate between the two types of theory in an
observational way. We may return to a more detailed consideration of this
interesting point elsewhere.

\section*{ Acknowledgments}

We would like to thank Colin Froggatt, Archil Kobakhidze, Rabi Mohapatra and
Holger Nielsen for useful discussions and comments. Z.K.\emph{\ }%
acknowledges financial support from Shota Rustaveli National Science
Foundation (grant \# YS-2016-81)

\end{document}